# Effect of Salt Concentration on Dielectric Properties of Li-Ion Conducting Blend Polymer Electrolytes

Anil Arya, A. L. Sharma*

Department of Physical Sciences, Central University of Punjab, Bathinda-151001, Punjab, INDIA

*Corresponding Author E-mail: alsharma@cup.edu.in

**Abstract**

In the present article, we have studied the effect of the salt concentration ($LiPF_6$) on transport properties and ion dynamics of blend solid polymer electrolyte (PEO-PAN) prepared by solution cast technique. Fourier transform infrared (FTIR) spectroscopy confirms the presence of microscopic interactions such as polymer-ion and ion-ion interaction evidenced by a change in peak area of anion stretching mode. The fraction of free anions and ion pairs obtained from the analysis of FTIR implies that both influence the ionic conductivity with different salt concentration. The complex dielectric permittivity, dielectric loss, complex conductivity have been analyzed and fitted in the entire frequency range (1 Hz-1 MHz) at room temperature. The addition of salt augments the dielectric constant and shift of relaxation peak in loss tangent plot toward high frequency indicates a decrease of relaxation time. We have implemented the Sigma representation ($\sigma''$ vs. $\sigma'$) for solid lithium ion conducting films which provide better insight toward understating of the dispersion region in Cole-Cole plot ($\varepsilon''$ vs. $\varepsilon'$) in lower frequency window. The dielectric strength, relaxation time and hopping frequency are in correlation with the conductivity which reveals the authenticity of results. Finally, the ion transport mechanism was proposed for getting the better understanding of the ion migration in the polymer matrix.

*Keywords: Blend Polymer Electrolyte, Electrical conductivity, Dielectric properties, relaxation time*

## 1. Introduction

At present, three complex global challenges are pollution, climate change and the lack of fossil fuels. The best alternative to overcome the issues mentioned earlier is by developing the renewable energy sector which will automatically reduce the dependency on the traditional or non-renewable energy resources [1-3]. In the modern world, rechargeable lithium-ion batteries (LIBs) are one of the most promising approaches toward a sustainable energy storage/conversion device since their introduction in 1991 by SONY Corporation. One and first advantage with the LIBs is that they are lighter in weight and produce high energy density. Their broad application range such as portable electronic devices, solar cells, supercapacitors and electrochromic devices makes them attractive as the future energy source. The electrolyte is a vital component of a battery system, and it plays the dual role, physical separation of electrodes as well as serves as an ion transporter during the cell operation. The overall performance of the battery is sharply affected by the performance of the electrolyte, and its role is essential in determining the performance parameters of the high energy density LIB [4-8]. Mostly, the liquid electrolyte and organic solvents were used as a traditional electrolyte which has some fundamental drawbacks such as leakage and safety hazards due to poor packaging. Also, to avoid the internal short circuit a separator (ceramic or polymer-based) is used between the electrodes. So, the much attention has been given to the solid polymer electrolyte (SPE) which overcomes the challenges mentioned above as well as permits us towards an economic, leakage free and flexible geometry. Another advantage with the SPEs is that it facilitates the fabrication of tiny batteries using thin electrolyte films and it automatically reduces both cost and weight [4, 9-12]. At the same time, solid polymer electrolytes are more stable and safer against existing systems.

Basically, the SPEs comprise of the dissolved lithium salt in a high molecular weight polymer host such as PEO and cation is active species which migrates via the coordinating interaction with the ether group of the polymer chain and the anion is supposed to be attached to the polymer backbone due to its bulky size. Therefore, for fulfilling the criteria of the advanced polymer electrolyte a number of SPE system based on various host polymer such as: poly(ethylene oxide) (PEO) [13-16] and related polymer based (e.g., polyacrylonitrile (PAN) [17-18], poly(vinylidene fluoride) (PVDF) [19], poly (vinyl chloride) (PVC) [20-21] , poly(methyl methacrylate) (PMMA) [22] have been synthesized. Out of different explored host polymer PEO ($-CH_2CH_2O-$) has gained a special interest due to its strong ability to dissolve the various alkali metal salt by the interaction of cation (Lewis Acid) with the electron-rich ether group (Lewis Base) [23]. It is a semi-crystalline polymer and exhibits low ionic conductivity at ambient conditions which restrict their use as an electrolyte in energy storage devices. As it is well known that the high amorphous content is required to promote the faster migration of the ion which leads to high ionic conductivity. To further improve the ion dynamics and transport properties polymers were modified by cross-linking and blending approaches etc, [24-28]. The polymer blending seems more promising and adaptable approach to obtain the enhanced properties in a controlled manner. It involves the desolvation of a crystalline polymer with an amorphous polymer as a partner to suppress the crystalline content using a common solvent. For blending, polyacrylonitrile (PAN) was chosen due to high thermal retardant nature, better mechanical and electrochemical properties. Also, the presence of electron rich polar nitrile (C≡N) group makes it interesting for blending with H of PEO. $LiPF_6$ has been chosen as conducting species due to its low lattice energy and smaller size of cation (w.r.t. anion) which results in faster ion migration from one electron rich site to another [28-31].

The present article is the extension of our previously published paper where, we discussed the structural, electrical and transport number analysis [32]. Since ion association/dissociation is the centralized issue and it directly affects the ionic conductivity which is further linked to the dielectric constant. The high dielectric constant supports the complete dissociation of the salt and hence the conductivity which is in direct proportion to the number of free charge carriers. So, this work is focused towards understanding the ion dynamics & dielectric relaxation in the case of the polymer electrolytes and is a great challenge to the researchers due to complex-ion dynamics in polymer electrolytes. So, it becomes crucial to investigate the dielectric properties such as dielectric constant, dielectric loss, ac conductivity, modulus and the relaxation time for getting insights into ion dynamics in the polymer salt matrix. It is important to note that, the low value of relaxation time results in the high ionic conductivity [33-35]. The frequency dependent dielectric parameters, ac conductivity, and modulus need to be investigated deeply for a better understanding of the ion transport mechanism in solid state ionic conductor (SSIC) [36]. At present, the impedance spectroscopy is frequently used for exploring the ion dynamics in the SSIC and data obtained from the impedance spectroscopy can be expressed in different representations such as dielectric permittivity, dielectric loss, tangent delta loss, Cole-Cole plot, ac conductivity, Sigma representation and electric modulus, etc., [37-38].

The present paper reports a systematic study of the effect of salt concentration on the dielectric properties and ion dynamics of prepared blend polymer electrolyte (BPE) system. Firstly, the impact of the salt concentration on the blend polymer electrolyte is analyzed by Fourier transform infrared spectroscopy. Thereafter, the complex permittivity is investigated with frequency as a variable parameter. We have used the Sigma representation for better exploring the dispersion region of the Cole-Cole plot and provides crucial information for examining the ion dynamics. All the dielectric parameters ($\varepsilon'$, $\varepsilon''$, tan $\delta$, $\sigma'$, $\sigma''$) were fitted in whole frequency window and evidences the substantial agreement between the experimental & fitted results. The fitting parameters such as dielectric strength, relaxation time were in absolute correlation with the conductivity value and the FTIR analysis which validated the investigated effect. An ion transport mechanism has been proposed for better visualization of the obtained experimental results.

## 2. Experimental

*2.1. Materials*

PEO (MW= $6\times10^5$ g/mol), PAN (MW=$1.5\times10^5$ g/mol) and $LiPF_6$ were purchased from Sigma-Aldrich. PEO, PAN, and $LiPF_6$ were dried under vacuum before use. N, N-Dimethylformamide (DMF), purchased from Sigma-Aldrich, was used as an aprotic solvent. The DMF is chosen due to its high dielectric constant (~36.71) and low viscosity (0.796) as well as good compatibility with both polymer and salt. DMF may be used as a common solvent for both the polymers (PEO & PAN). The blend polymer PAN is unassociated in DMF as compared to DMSO, DMAA [3, 9] as DMF is best known universal solvent for polymers.

*2.2. Preparation of Solid polymer electrolytes*

The amount of PEO (0.5 g) and PAN (0.5 g) was kept fixed for all solid polymer electrolytes, and concentration of salt was varied as Ö/Li=8, 10, 14, 16, 18, 20 and is calculated as

$$\frac{\text{Ö}}{\text{Li}^+} = \frac{\text{No. of monomer unit in half a gram of PEO}}{\text{No. of LiPF6 molecular in half a gram of salt}} \times \frac{\text{wt. of PEO taken}}{\text{wt. of salt taken}} \quad (1)$$

First, the appropriate amount of the PEO was dissolved in DMF at room temperature and stirred on a magnetic stirrer. Then an appropriate amount of PAN was added in the polymer solution and again stirred on a magnetic stirrer followed by adding salt in a stoichiometric ratio of salt (Ö/Li) in the polymer blend matrix.

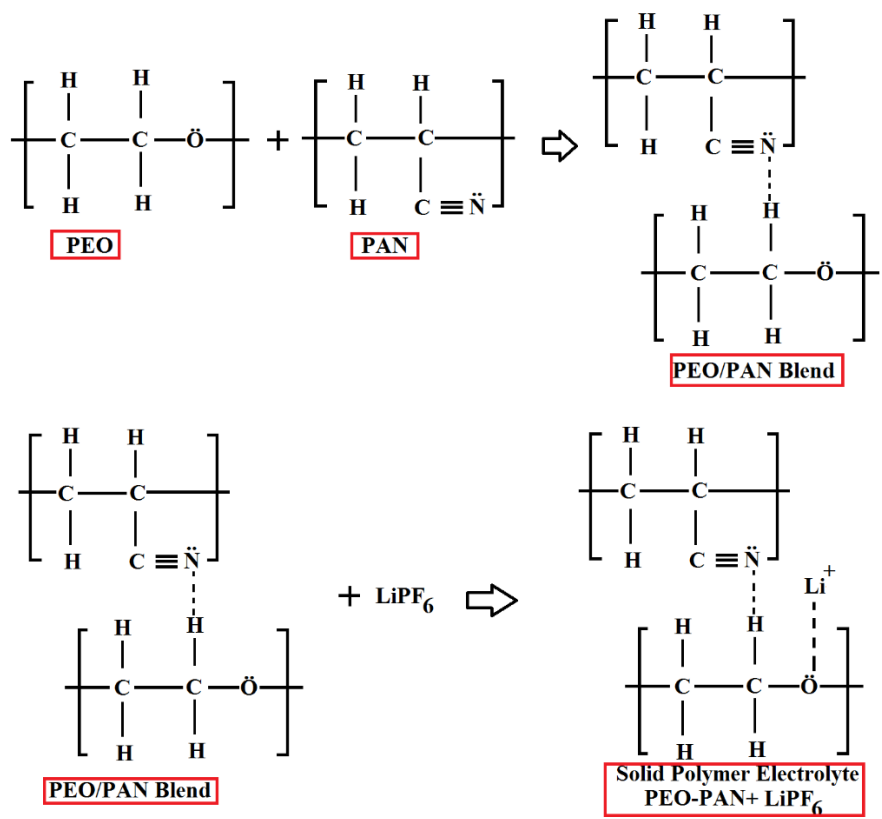

**Figure 1**. An interaction mechanism between the host polymer/blend polymer and the lithium salt.

The whole mixture was again stirred for 10 hours, and the obtained solution was cast on the glass Petri dishes and kept for a few days for evaporation of the solvent at room temperature. Further to remove the residue of solvent it was dried for 24 hours in a vacuum oven. The free-standing solid polymer electrolyte films were obtained by peeling off the glass petri dish and kept in a vacuum desiccator for further characterization. Figure 1 depicts the interaction occurring in the two polymers, which confirms the formation of the blend polymer and lithium (cation) coordination with the ether group of the PEO. Here two interaction possibilities arise in the blend formation one is proton of PEO with N of PAN and proton of PAN with O of PEO.

*2.3. Characterization and measurements*

Fourier transform infrared (FTIR) spectroscopy of these samples was recorded to check the complex formation and presence of polymer-ion and ion-ion interaction in the wavenumber region 600 cm$^{-1}$ to 3000 cm$^{-1}$ with (Bruker, Tensor 27 NEXUS–870). For the electrical measurements, the polymer electrolytes films were sandwiched between two stainless steel (SS) blocking electrodes in a configuration; SS|SPE|SS and ac signal of ~10 mV is applied with (Model: CHI 760, USA) across it. The impedance data was further transformed into the dielectric parameters such as $\varepsilon'$, $\varepsilon''$, $\tan\delta$, $\sigma'$, $\sigma''$, $M'$ and $M''$.

3. **Results and discussions**

3.1. *Fourier Transform Infrared Spectroscopy (FTIR) Analysis*

Fourier Transform Infrared Spectroscopy (FTIR) was performed for evidencing the cation-polymer complexation concerning chemical bonding, amount of free anion/ion pair and to verify the ion-ion interaction in the polymer electrolyte. The FTIR spectrum of the pure polymer blend (PEO-PAN) and with different salt composition (Ö/Li = 14, 16, 18 and 20) is shown in Figure 2. The vibrational peak at 954 cm$^{-1}$ corresponds to the C-O stretching vibration, and the dotted circular arc guides spectral pattern for the same. The specific vibrational mode at 1103 cm$^{-1}$ was assigned to the symmetric and asymmetric C-O-C stretching of host polymer (PEO). The vibration bands corresponding to symmetric twisting, asymmetric twisting, and bending of CH$_2$ are located at 1236 cm$^{-1}$, 1282 cm$^{-1}$, and 1352 cm$^{-1}$ respectively for the polymer host (PEO) [39-40]. The peak observed at 2245 cm$^{-1}$ is for the nitrile group of PAN and it confirms that the blend formation occurs. Further on complexation with the salt the polymer blend exhibit changes in the peak position and shape of a C-O-C mode of PEO, while there was no change in the nitrile group associated with PAN. This suggests that the preferable site for cation coordination is ether group of PEO which is more electronegative than the nitrile group. The variations in the intensity, shape, and location of the C–O–C symmetric/asymmetric stretching mode were associated with the interactions between the polymer host (PEO) and cation (Li$^+$) of LiPF$_6$. This electron rich ether group of PEO was responsible for the ion conduction and provides the coordinating sites to the cation for migration, while PAN supports the backbone of the matrix.

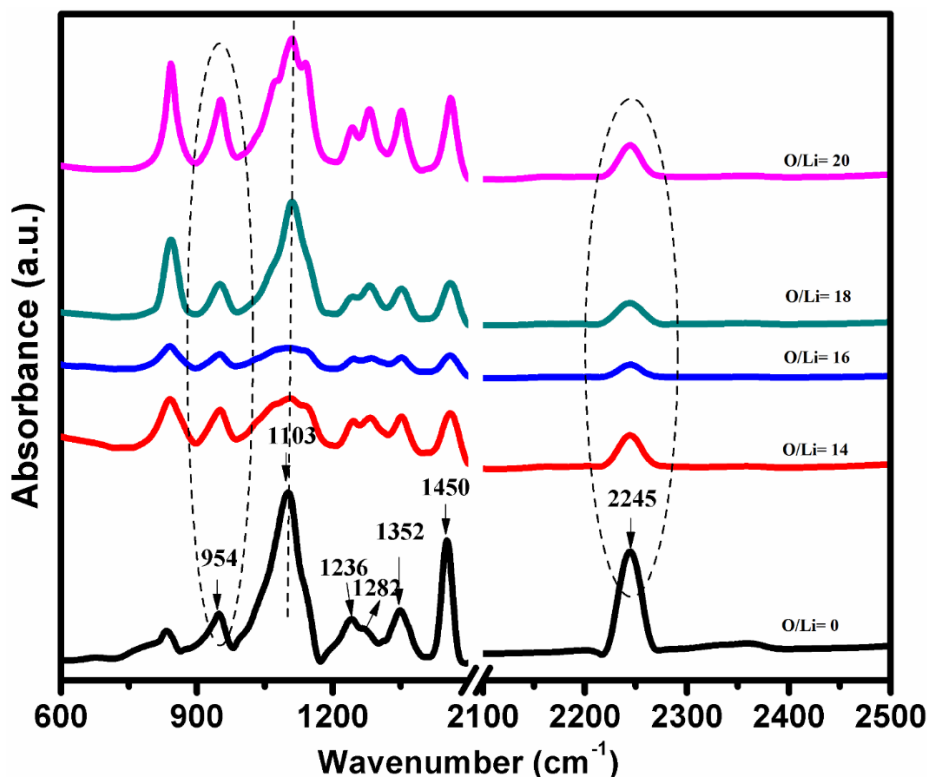

**Figure 2**. FTIR spectrum for the polymer blend PEO/PAN+LiPF$_6$ [**a** O/Li=0, **b** O/Li=14, **c** O/Li=16, **d** O/Li=18, and **e** O/Li=20] based solid polymer electrolyte.

Now, the signature region evidencing polymer-ion interaction is depicted by the peak located near 1100 cm$^{-1}$ corresponding to C-O-C stretching of host polymer (PEO). When the salt is added to the polymer blend the peak position get shifted slightly (2-4 cm$^{-1}$) toward the lower wavenumber side due to complexation of the Li$^+$ with ether group (-O-) of the host polymer. Further addition of the salt changes the peak area as well as the intensity of the peak which indicates that the cation get coordinated with the electron-rich ether group and complexation was formed as shown in recent section by schematic diagram in figure 1. Further, the decrease of peak area indicates the enhancement in amorphous content (figure 2 c) and evidences the complete dissociation of the salt which reflects the availability of more free charge carriers [41].

To study the ion-ion interaction, the spectral pattern of the $PF_6^-$ mode ($v_3(t_{1u})$ mode) is demonstrated in figure 3 (a-c) in wavenumber region 800-900 cm$^{-1}$. It is noticeable that the presence of asymmetry in the peak is owing to the simultaneous presence of more than two components, i.e. free ion, ion pairs. So, the deconvolution of the $PF_6^-$ was performed using the PeakFit (V4) software and it results in splitting into two distinct peaks located at 837 cm$^{-1}$ and 858 cm$^{-1}$. The peak at lower wavenumber is assigned to free $PF_6^-$ anions which do not directly interact with the lithium cations and at higher wavenumber is due to $Li^+ -- -PF_6^-$ ion pairs respectively [42-43]. The amount of free ion area ($PF_6^-$) and ion pair area ($Li^+PF_6^-$) was estimated from the De-convoluted pattern and a dual y-axis plot represents the both (free anion and ion pair) contribution against the salt concentration (Ö/Li) in Figure 3 d. The highest fraction of free ion area was observed for the critical concentration Ö/Li=16. The improvement in the fraction of free anion area must encourage faster ion dynamics and transport properties of the present system. Therefore the complex impedance analysis has been performed in the next section.

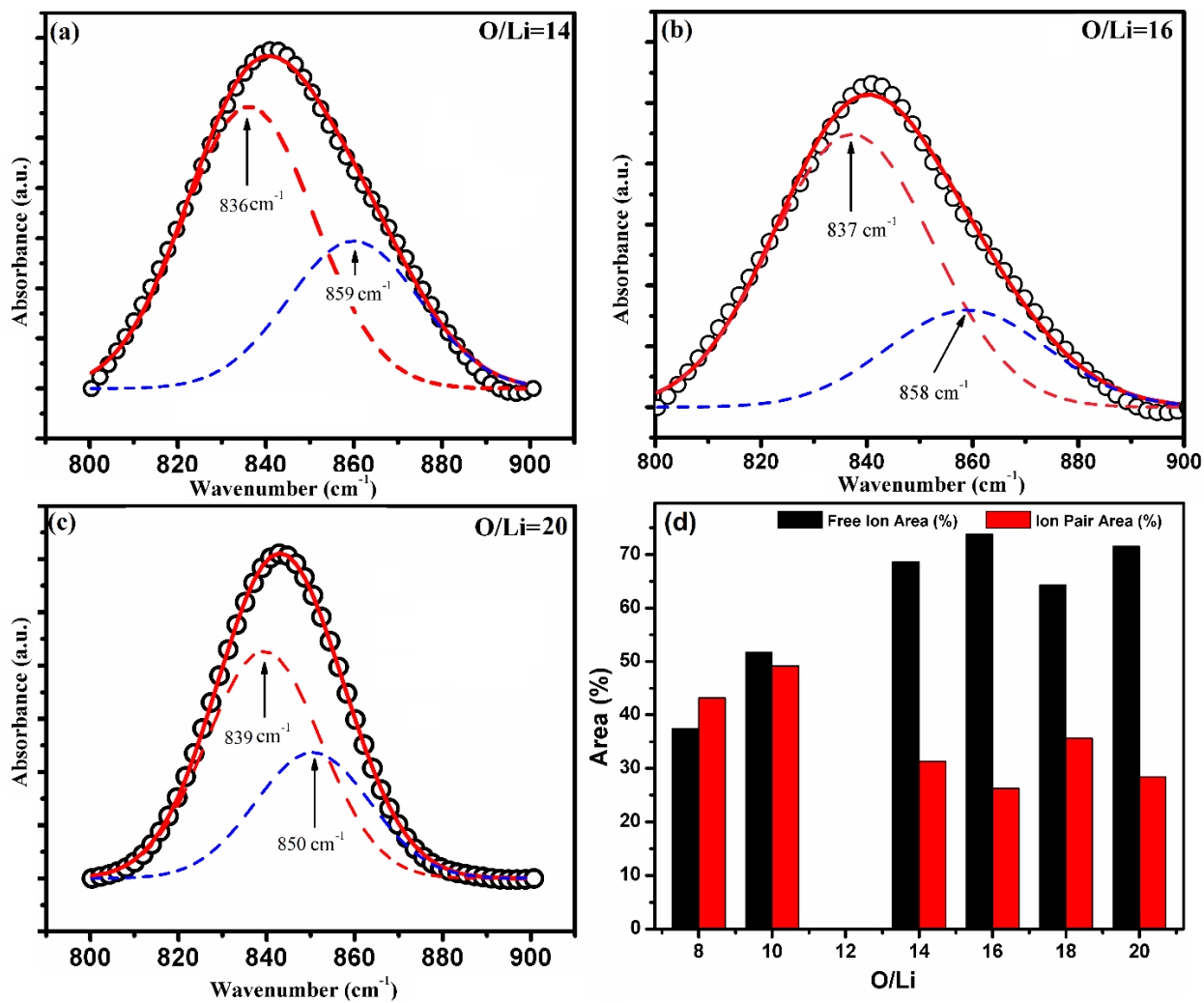

**Figure 3.** Deconvolution of the $PF_6^-$ vibration mode in the wavenumber range 800 cm$^{-1}$ to 900 cm$^{-1}$ containing different salt concentration, (a)O/Li=14, (b)O/Li=16, (c) O/Li=18 and (d) Plot of free ion area and ion pair area against the salt concentration.

*3.2. Complex Impedance Analysis*

The impedance analysis of the prepared solid polymer electrolyte film was carried out by sandwiching the polymer electrolyte between the two stainless steel (SS) electrodes in the composition SS|SPE|SS. The SS electrodes play the role of blocking electrodes for Li$^+$ on the application of the electric field. Figure 4 shows the complex impedance plots ($Z''$ vs. $Z'$) and their fitting results are indicated by solid line. For better clarity and comparison of the different impedance plots in the single plot, the log-log presentation of these complex impedance plot was represented. Here, it was observed that the semicircular arcs were distorted, but logarithmic plots were superior in various aspects as discussed by Jonscher [44-45]. Two semi-circular arcs were observed for the polymer blend, one corresponds to the semicircular arc and another corresponds to the spike at low-the frequency side in the linear impedance plot. The first semicircular arc get suppressed with the addition of the salt. Further deep analysis evidence that the first arc shift from right to left in the graphical representation and indicates decrease of associated impedance. The minima in the arc associated with the y-axis gives the $Z''$ and for the corresponding minimum value of $Z''$ the value of $Z'$ (on x-axis) is bulk resistance ($R_b$). The reduction in the semicircular arc with the addition of salt may be attributed to the formation of space charge at the electrode-electrolyte interface region, entitled as double layer capacitive effect and total conductivity is dominated by ion conduction only [46].

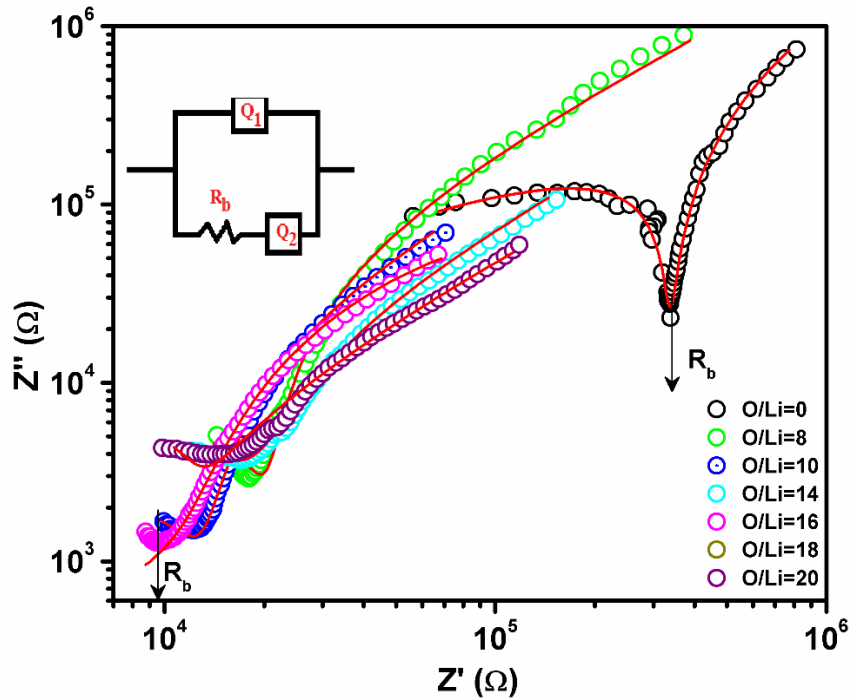

**Figure 4.** Log-log plots of complex impedance ($Z''$ vs. $Z'$) for different salt concentration. Inset shows the fitted equivalent circuit.

For analyzing the impedance data in detail the equivalent circuit (shown in graph as inset) was fitted which consists of the constant phase element ($Q_1$) in parallel with the series combination of the resistance ($R_b$) and another constant phase element ($Q_2$). The solid lines in the graph displays best fit for the corresponding impedance plot and equivalent circuit confirms the absolute agreement between the experimental and fitted results. It may be concluded in terms of change in pattern from the graph that addition of salt strongly influences the electrochemical properties and shift of dip in the plot toward the left side indicates the reduction of bulk resistance and hence the increase of the ionic conductivity .

### 3.3. Dielectric Spectroscopy

#### 3.3.1. Cole-Cole Plot

The Cole-Cole plot ($\varepsilon''$ $vs.$ $\varepsilon'$ ) is proposed to study the single relaxation processes and is helpful in interoperation of the dielectric measurements. This plot comprises of the imaginary part of dielectric permittivity (dielectric loss) against the real part of dielectric permittivity (dielectric constant) with frequency as the parameter (figure 5). The plot for the materials obeying Debye equation is perfect semicircle with two points of intersection on real axis attributed to the zero loss and expressed by the equation 2 [47].

$$\left(\varepsilon' - \frac{\varepsilon_s + \varepsilon_\infty}{2}\right)^2 + (\varepsilon'')^2 = \frac{(\varepsilon_s - \varepsilon_\infty)^2}{2} \qquad (2)$$

First one corresponds to the dielectric constant of infinite frequency ($\epsilon_\infty$) (left part in graph) and second is static dielectric constant ($\epsilon_s$) (right part in graph) [48].

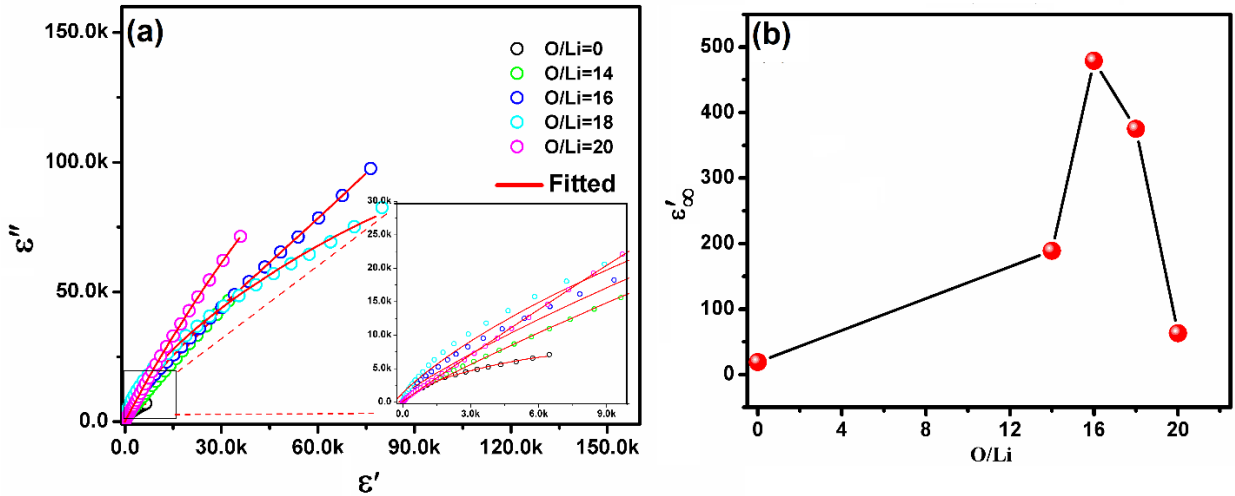

**Figure 5.** (a) Cole-Cole plot for the polymer blend PEO/PAN+LiPF$_6$ and (b) Variation of infinite dielectric constant against the salt concentration based solid polymer electrolyte.

It is also evidenced from the plot that the diameter of dotted semi-circle increases on the addition of the salt in the BPE that is an indication of the increased number of ions which contributes to ion conduction from one coordinating site to another. In the plot, a semicircular nature appears and followed by a dispersion region toward the high-frequency window. This dispersion region is an indication of high loss in the polymer electrolyte. Also, the different value of the $\epsilon_\infty$ for all polymer electrolytes evidence the presence of active dipoles in the investigated system. The high value of the dielectric constant for the O/Li=16 is attributed to the availability of more free number of charge carriers due to better salt dissociation (Figure 5 b) and if this value remains same that evidences that dipoles are stationary dipoles [49-50]. This is also in good agreement with the impedance study and the free ion area as obtained from the De-convolution of FTIR. A more detailed investigation is required here to completely analyze the role of salt in the ion dynamics, so the next section provides batter glimpse of the dielectric analysis.

### 3.3.2. Real part of Dielectric Permittivity

The frequency dependent dielectric parameters dielectric constant and dielectric loss need to be studied for better understanding thr role of the salt in enhancing the ionic conductivity. Former one basically defines the polarizing ability of material on application of the field while the latter one is an indication of energy loss in aligning the molecular dipoles in the field direction [51-52]. The complex dielectric permittivity ($\varepsilon^*$) is expressed by equation 3:

$$\varepsilon^* = \varepsilon' - j\varepsilon''; \quad \varepsilon' = \frac{-Z''}{\omega C_o(Z'^2+Z''^2)} \text{ and } \varepsilon'' = \frac{Z'}{\omega C_o(Z'^2+Z''^2)} \quad (3)$$

Where, where $\varepsilon'$ and $\varepsilon''$ are the real and imaginary parts of the dielectric permittivity respectively. The real and imaginary parts of dielectric constant can obtained by separating above equation and can be given as equation 5 a & b [51, 53, 54]

$$\begin{cases} \varepsilon' = \epsilon_\infty + \dfrac{\Delta\varepsilon\left(1 + x^\alpha \cos\frac{\alpha\pi}{2}\right)}{1 + 2x^\alpha \cos\frac{\alpha\pi}{2} + x^{2\alpha}} & (5\ a) \\ \varepsilon'' = \Delta\varepsilon \dfrac{x^\alpha \sin\frac{\alpha\pi}{2}}{1 + 2x^\alpha \cos\frac{\alpha\pi}{2} + x^{2\alpha}} & (5\ b) \end{cases}$$

Here, $\varepsilon_s$ is static dielectric constant ($x \to 0$), $\varepsilon_\infty$ is dynamic dielectric constant ($x \to \infty$), $x = \omega\tau$; $\omega$ is angular frequency of applied field and $\tau$ is average Debye relaxation time. Here, $\alpha$ is distribution exponent of material sample. The fitted parameters are shown in the Table 1 at RT. From Table 1 it is observed that decrease in value of $\alpha$ is direct indication of more distributed relaxation time.

Table 1. The fitted $\varepsilon'$ ($\varepsilon_\infty$, $\Delta\varepsilon$, $\tau_{\varepsilon'}$, $\tau_m$, $\alpha$) and $\varepsilon''$ ($\Delta\varepsilon$, $\tau_{\varepsilon''}$, $\alpha$) parameters at room temperature.

| Sample Code | $\varepsilon'$ | | | | | $\varepsilon''$ | | |
|---|---|---|---|---|---|---|---|---|
| | $\varepsilon_\infty$ | $\Delta\varepsilon(\times 10^3)$ | $\tau_{\varepsilon'}(\mu s)$ | $\tau_M(\mu s)$ | $\alpha$ | $\Delta\varepsilon(\times 10^3)$ | $\tau_{\varepsilon''}(\mu s)$ | $\alpha$ |
| O/Li=0 | 19 | 10.27 | 1.69 | 1.12 | 0.83 | 19.93 | 1.82 | 0.78 |
| O/Li=14 | 189 | 91.16 | 0.57 | 0.37 | 0.64 | 401.70 | 1.08 | 0.63 |
| O/Li=16 | 479 | 150.57 | 0.12 | 0.07 | 0.68 | 564.53 | 0.33 | 0.63 |
| O/Li=18 | 375 | 121.89 | 0.61 | 0.40 | 0.75 | 280.38 | 0.17 | 0.65 |
| O/Li=20 | 63 | 137.02 | 0.99 | 0.65 | 0.72 | 715.65 | 0.64 | 0.72 |

The frequency dependent real part of the dielectric permittivity was shown in Figure 6 and all plots demonstrate a decrease of dielectric constant with an increase in the frequency. The solid line in the graph displays the best fitting result. The addition of the salt in the polymer blend matrix enhances the dielectric constant and is directly linked to the number of free charge carriers. The high value of the dielectric constant at the low-frequency window is due to electrode polarization event which remains associated with the accumulation of the ions and evidence the complete dissociation of the salt. This nature also confirms the non-Debye dependence [55-56]. The blocking electrodes prevent the ion migration to the external circuit, and this results in the accumulation of ions on the opposite electrodes, termed as polarization. Also in the low-frequency window, the ion pairs remain in the immobilize state which hinders the long-range motion and results in the high value of the dielectric constant due to sufficient time [57].

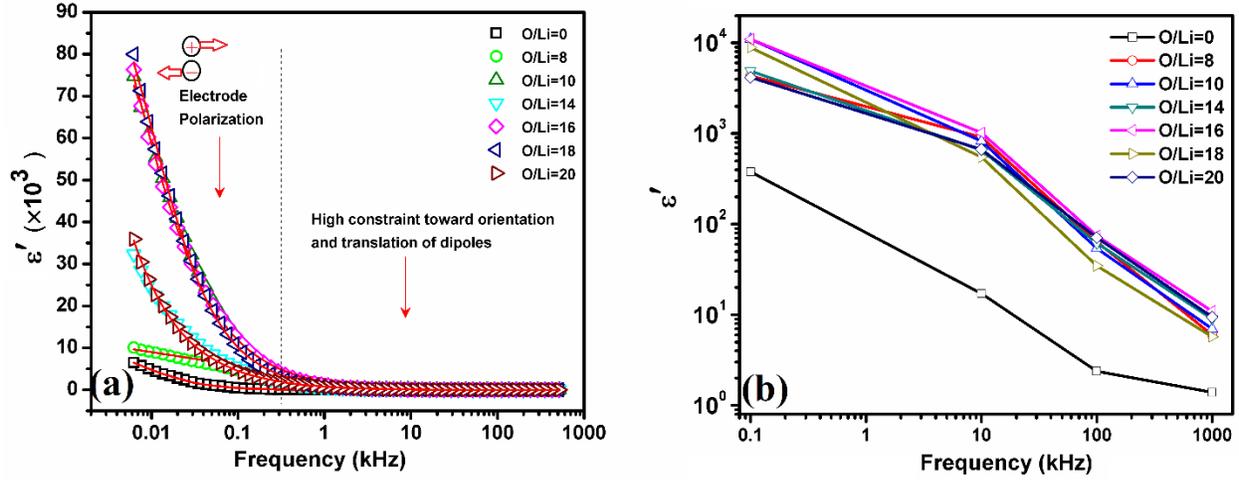

**Figure 6. (a)** Frequency dependence of the dielectric constant ($\varepsilon'$) for polymer blend PEO/PAN+LiPF$_6$ and **(b)** Variation of dielectric constant with frequency for different salt concentration. Solid lines are the best fit to the experimental data.

Now, the high-frequency window the decrease of the dielectric constant is attributed to the dominance of the relaxation process. Here, the rapid change in the direction of the field makes ions incapable of responding to the applied field due to lack of inadequate time for rotation/translation of dipoles. So, now due to insufficient time ions are unable to accumulate at the electrodes and dielectric permittivity decreases [58]. This region may be treated as the frequency independent region and indicates the failure of dipoles to follow the field direction. Figure 6 b shows the variation of the dielectric permittivity at different frequencies and dielectric constant decreases with the increase of the frequency. The high value of dielectric constant was obtained for the optimized system. This is in correlation with the ionic conductivity value which directly depends on the number of free charge carriers $\sigma_i = \sum_i q_i n_i \mu_i$ .

From the Table 1, it is concluded that the dielectric strength ($\Delta\varepsilon = \varepsilon_s - \varepsilon_\infty$) increases with the addition of salt while relaxation (both average and molecular) time decreases. The former might be attributed to the complete dissociation of the salt via the polymer-ion interaction, and high value of dielectric constant achieved for this confirms the same. Now, for measuring the ionic mobility, the relaxation time is important. Here, the decrease of relaxation time means that the ion jumps faster from one coordinating site to another via the polymer host and it infers that the polymer chain mobility increase on addition of salt. As, the ionic conductivity is inversely proportional to the relaxation time. So decreased value of relaxation times results in the highest ionic conductivity and is in agreement with the impedance study as explained in recent section Furthermore, we have calculated the molecular relaxation time using the equation 6.

$$\tau_m = \left[\frac{(2\varepsilon_s + \varepsilon_\infty)}{3\varepsilon_s}\right] \times \tau_{\varepsilon'} \qquad (6)$$

The molecular relaxation time also follows the same trend and is in absolute correlation with the aforementioned impedance and the FTIR analysis. The decreased molecular relaxation time is due to the increased number of free charge carriers owing to the fast segmental motion of polymer chains [50]. This high value of the dielectric constant for the O/Li=16 system also in good correlation with the impedance analysis. Further, the FTIR deconvolution for the same concentration provides release of the more free number of the charge carriers and supports faster ion migration (Figure 4).

### 3.3.3. Imaginary part of dielectric permittivity

Figure 7 showed the plot of the imaginary part of the dielectric permittivity against the frequency and termed as a dielectric loss ($\varepsilon''$). It indicates the amount of energy required to align the dipoles in the direction of the field. The solid lines represent the absolute fit to the experimental data using equation 4 b.

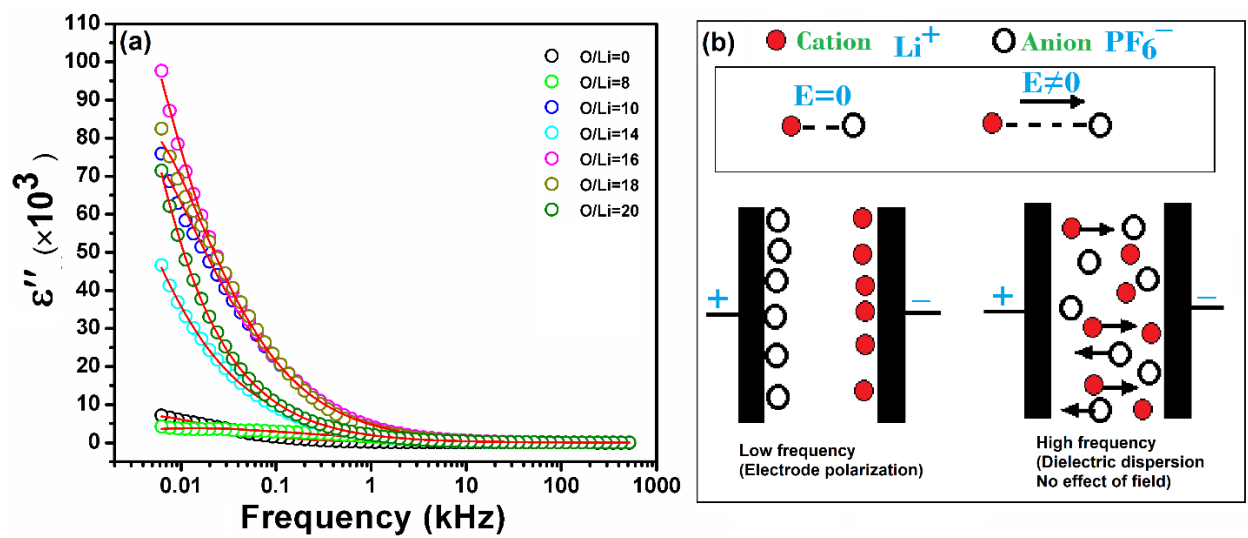

**Figure 7.** Frequency dependence of the dielectric loss ($\varepsilon''$) for polymer blend PEO/PAN+LiPF$_6$ [a, O/Li=0; b, O/Li=14; c, O/Li=16; d, O/Li=18; and e, O/Li=20] based solid polymer electrolyte and (f) double layer formation at low frequency. Solid lines are the best fit to the experimental data.

As the polymer electrolyte system comprises of ion-ion and polymer-ion interaction which results in almost complete dissociation of the salt. So, ions are the active species which response to the applied external electric field. During the periodic reversal of the field, the polymer electrolyte system follows a three-step process before reversing the direction. In the first step just when the field direction changes, at that moment de-acceleration of the ions occurs. Then in the second step ion comes in the stationary position and stays for a nonzero time there. Finally, in the third step ion is again accelerated in the reverse direction. The three-step process results in the heating of the dielectric polymeric system and this internal heat is called dielectric loss [59]. The value of the dielectric loss approaches to zero for the zero relaxation time ($\varepsilon'' = 0$ for $\omega\tau = 0$). This is also verified from the fitted parameters as shown in Table 1. The lowering in value of relaxation time as compared to the polymer blend system indicates the faster ion migration. The absence of any relaxation peak may be attributed to the masking effect, in which relaxation behavior get masked due to dominant electrode polarization (EP) effect. So, for the better understaing we approach for the alternative such as electric modulus and ac conductivity analysis, as discussed in the coming sections.

*3.4. Dielectric loss tangent*

The loss tangent vs. frequency plot is known as loss tangent plot as shown in Figure 8. The loss tangent (tan δ) is the ratio of imaginary part of permittivity to the real part of permittivity or ratio of energy loss to energy stored. Initially, the increase in the loss with an increase in the frequency is observed and the maxima at the particular frequency (where $\omega\tau = 1$) is followed by the decrease in high frequency. This plot can be divided into the three regions for a better understanding of the variation of ac conductivity with frequency, first is a low-frequency region, second is moderate frequency region, and the third one is a high-frequency region (Figure 8).

All graph shows a single relaxation peak which indicates ionic conduction in the present system. In the low-frequency region increase of the loss associated with the dominance of the Ohmic component than the capacitive element. While the presence of the maxima is observed only at a single frequency when the perfect matching between the frequency of electric field and frequency of molecule rotation occurs. This resonance leads to the maximum power transfer to the dipoles in the system and hence the maximum heat [60]. Now, in the high-frequency window, the capacitive component becomes dominant. Here, the Ohmic part becomes frequency independent, while the capacitive part grows with the frequency [61-62].

Now, the effect of salt concentration is investigated. It is noticeable from the figure that the relaxation peak shifts toward the high-frequency side which indicates the faster ion dynamics from one coordinating site to another due to a decrease of relaxation time. To verify the above said results and calculation of relaxation time the fitting of the tangent delta plot was performed with the equation 7 proposed by the Debye [53].

$$\tan \delta = \frac{(r-1)}{r + x^2} x \quad (7)$$

Where r is the relaxation ratio ($\varepsilon_s/\varepsilon_\infty$), $\varepsilon_s$ is static dielectric constant ($x \rightarrow 0$), $\varepsilon_\infty$ is dielectric constant ($x \rightarrow \infty$), $x = \omega\tau$; $\omega$ is the angular frequency of applied field and $\tau$ is Debye relaxation time (reciprocal of jump frequency in the absence of external electric field). But, during the fitting and general analysis, it appears that the fitting is absolute only in the high frequency region. Since the Debye equation is valid only for the single dipole relaxation and non-interacting system therefore same theoretical equation may not be appropriate. Hence in, the present investigated system the presence of the multi-type dipole polarization is owing to the complex system. Also, the broad relaxation peak suggests us to modify the Debye equation 7 for perfect fitting in whole frequency window which certainly overcomes the issue of poor low frequency fit. So, to meet the experimental needs in the present system one parameter (α) is added as the power law exponent with value lying between $0 \leq \alpha \leq 1$ as expressed by the equation 8. Many Earlier reports also give the information about the empirical modification done by adding some parameter, e.g. one parameter was used in Cole-Cole, Davidson-Cole, Williams-Wats, and two parameters in Havriliak-Negami fluctuations [63].

$$\tan \delta = \left(\frac{(r-1)}{r + x^2} x\right)^\alpha \quad (8)$$

This equation was fitted to the tangent delta plot and fits well in the whole frequency window. The fitted parameters obtained by the best fitting (represented by the solid line) are summarized in Table 5. The presence of a small deviation in the experimental results may be due to greater electrode polarization/diffusion of ions towards the electrodes. As no electrode is perfectly blocking, so it may result in the low-frequency substantial deviation in some system [64-65].

Table 2. The fitted tangent delta loss parameters (r, τ, α. $\tau_{\tan \delta}$, $\tau_m$) at room temperature.

| Sample Code | r (×10$^6$) | τ | α | $\tau_{\tan\delta}$ (μs) | $\tau_m$ (μs) |
|---|---|---|---|---|---|
| O/Li=0 | 3 | 0.08 | 0.54 | 419 | 2.113 |
| O/Li=14 | 29 | 0.22 | 0.19 | 53 | 0.007 |
| O/Li=16 | 156 | 0.23 | 0.24 | 18 | 0.001 |
| O/Li=18 | 5 | 0.15 | 0.34 | 66 | 0.002 |
| O/Li=20 | 17 | 0.25 | 0.18 | 62 | 0.015 |

An interesting point to be noted is that α =1 leads to the Debye equation. The physical significance of α was not yet been worked out but to be done in future and it will provide crucial aspects which justify the proposed equation. To further investigate the dielectric analysis fully, in the next section, suppressed features of Cole-Cole plot at high frequencies were explored with a new representation in case of solid state ionic conductors (SSICs).

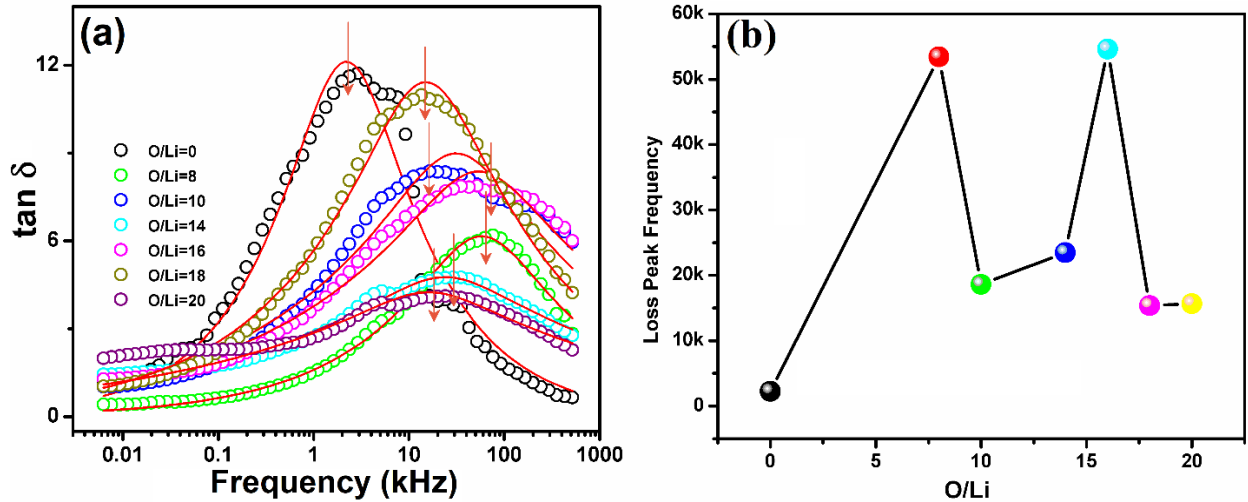

**Figure 8.** (a) Frequency dependence of the tangent delta loss (tan δ) for polymer blend PEO/PAN+LiPF$_6$ based solid polymer electrolyte and (b) Variation of loss peak frequency against the salt concentration. Solid lines are the best fit to the experimental data.

*3.5. Sigma Representation*

As in the previous section, the dielectric behavior was explained using the Cole-Cole plot ($\varepsilon''$ vs. $\varepsilon'$) and is helpful for the material following the Debye equations. But, at low frequency dispersion in the Cole-Cole plot was observed that may be due to the presence of dc conductivity component. So, for the better understanding of the ion dynamics, it becomes important to analyze the dispersion behavior along with the whole frequency region. Therefore, an important approach is known as Sigma representation used and it provided us some important aspects which will change the understanding of the ion dynamics in polymer electrolytes [66]. Another important finding is that the high frequency features which were suppressed in the Cole-Cole plot can be easily notified in Sigma representation because both the components of the complex conductivity involves multiplication of the frequency term. Sigma representation comprises of the imaginary part of complex conductivity ($\sigma''$) against the real part of complex conductivity ($\sigma'$) with frequency as the parameter. This plot comprise of a semicircle with center on the real axis and intersects at $\sigma_o$, $\sigma_\infty$ respectively (Inset of figure 9 ). The solid red line in the plot is absolute ffit for the sigma representation plot. As the complex electrical conductivity can be written using following expression (equation 9);

$$\begin{cases} \sigma(\omega) = \sigma' + i\sigma'' & (9\ a) \\ \sigma_\infty = \sigma_o + \dfrac{\varepsilon_v(\varepsilon_o - \varepsilon_\infty)}{\tau} = \sigma_o + \delta & (9\ b) \\ \sigma_{ac} = \sigma' = \omega\varepsilon_v\varepsilon'' \text{ and } \sigma_{dc} = \sigma'' = \omega\varepsilon_v(\varepsilon' - \varepsilon_\infty) = \omega\varepsilon_v\varepsilon' & (9\ c) \\ r = \dfrac{\delta}{2} = \dfrac{\varepsilon_v(\varepsilon_o - \varepsilon_\infty)}{2\tau} & (9\ d) \end{cases}$$

Here, $\sigma'$ is the real part of conductivity, $\sigma''$ is the imaginary part of conductivity, $\omega$ is the angular frequency, 'r' is the radius of the semicircle. And, when $\sigma'' = 0$ then low frequency x-intercept gives dc conductivity ($\sigma_o$) and high frequency x-intercept gives $\sigma_\infty$. The radius of the semicircle (r) is inversely proportional to the relaxation time ($\tau$) and is a crucial parameter since it allows us to know whether the ion dynamics is faster or slower on the addition of salt. The plot was displayed in the figure 9 and all patterns shows clear semicircle nature which confirms the Debye type behavior. The intercept on the real axis provides two conductivity parameters one is $\sigma_o$ and another is $\sigma_\infty$ respectively. The parameter obtained from the plot are listed in the Table 3.

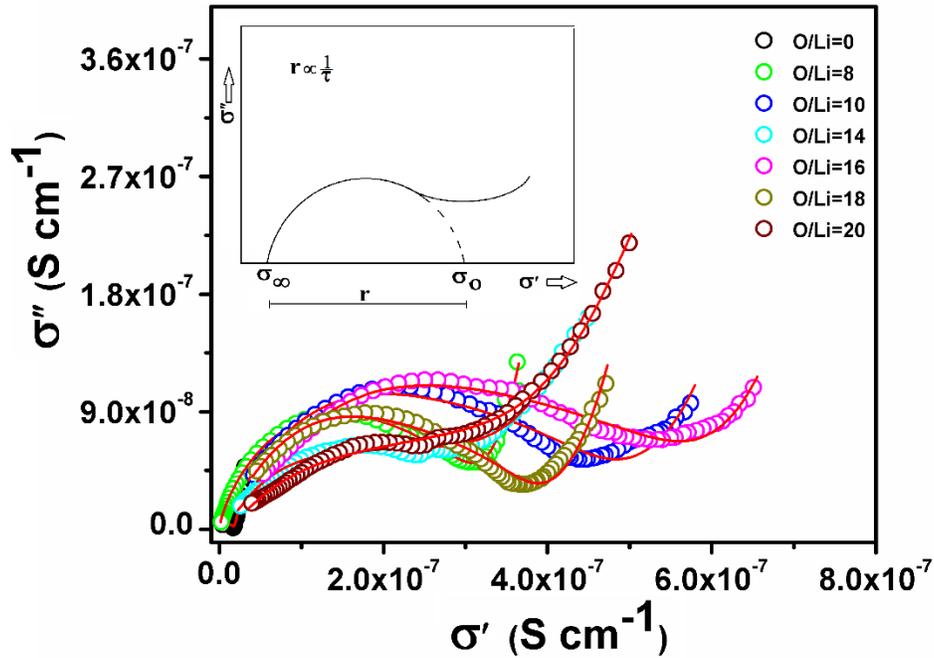

**Figure 9**. $\sigma''$ vs. $\sigma'$ plot for polymer blend PEO/PAN+LiPF$_6$ blend solid polymer electrolyte.
All the graphs show a semicircular nature with a tilted spike at low frequency toward the right side of the figure. The addition of salt changes the pattern and depressed semicircle with increased radius were obtained. Since the radius is inversely proportional to the relaxation time so it directly provides the evidence of decreased relaxation time hence improved conductivity. It may be noted from the figure 9 that the radius was maximum for the polymer electrolyte with O/Li=16 concentration and reflects the highest ionic conductivity at this salt concentration (Eq 9 d). This analysis provides the substantial evidence of the highly ionic conductive system for the same concentration (O/Li=16) and in correlation with the impedance and the FTIR deconvolution results. The following section comprises the real and imaginary part of the complex conductivity and also experimental data is fitted with the corresponding equations.

### 3.5.1. The real part of Complex Conductivity

The ac electrical measurements (ac conductivity) of all polymer electrolyte has been obtained using the equation 10:

$$\sigma' = \sigma_{ac} = \omega \varepsilon_o \varepsilon'' = \omega \varepsilon_o \varepsilon' \tan \delta \quad (10)$$

Where $\omega$ is the angular frequency, $\varepsilon_o$ is the dielectric permittivity of the free space and $\varepsilon''$ represents the dielectric loss. Figure 10 shows that variation of the ac conductivity against the frequency. The plot consists of three distinct regions depending upon the frequency. First is the low-frequency region with a sharp rise in the conductivity due to electrode polarization, followed by frequency independent plateau region at an intermediate frequency (corresponds to dc conductivity) and, high-frequency dispersive region owing to the fast reversal of the field. All figure obviously indicates that at low frequency there was a decrease in the conductivity value and may be due to the dominance of the electrode polarization (EP) effect. The constant region in the intermediate frequency window may be attributed to the long-range conduction of the charged carriers and dc conductivity is extracted from it. The high-frequency dispersion region is due to the short range ion transport (hopping) associated with ac conductivity. Blend polymer electrolyte (BPE) and polymer salt system shows all the three regions. In the polymer blend, all three regions appear apparently while on the addition of the salt high-frequency region shifts toward the right, known as dispersion region and falls outside the measured frequency range [67-68]. The high-frequency region is obtained in the polymer blend and small appearance in other polymer electrolyte system follows the fundamental Jonscher's universal power law (JPL) which is true characteristics of an ionic conductor, (equation 11 a & b)

$$\begin{cases} \sigma_{ac} = \sigma_{dc}(1 + (\omega/\omega_h)^n) & (11\ a) \\ \quad \text{and} \\ \sigma_{ac} = 2\sigma_{dc} \text{ when } \omega = \omega_h & (11\ b) \end{cases}$$

$\sigma_{ac}$ and $\sigma_{dc}$ are the ac and dc conductivities, A and n are the frequency independent Arrhenius constant and the power law exponent with value $0 < n < 1$. $\omega_h$ is hopping frequency, and at this particular frequency, ac conductivity becomes double of dc conductivity [69-71]. Although, JPL provides sufficient information about the hopping mechanism, it agrees sound only at higher frequencies and fails at lower frequencies due to electrode polarization effect. So, an alternative approach was used to analyze entire frequency window was used in the present investigated system which also includes the contribution from the EP effect and provides us more information which will enhance the understating of the ion dynamics and ion transport phenomena in polymer electrolytes [72]. The effective complex conductivity can be written as

$$\sigma^*_{eff} = \left(\frac{1}{\sigma_b} + \frac{1}{i\omega C_{dl}}\right)^{-1} + i\omega C_b \quad (12)$$

Now, considering the equation 13, the real and imaginary part of the conductivity can be written as

$$\sigma'(\omega) = \frac{\sigma_b^2 C_{dl} \omega^\alpha \cos\left(\frac{\alpha \pi}{2}\right) + \sigma_b (C_{dl}\omega^\alpha)^2}{\sigma_b^2 + 2\sigma_b C_{dl}\omega^\alpha \cos\left(\frac{\alpha \pi}{2}\right) + (C_{dl}\omega^\alpha)^2} \quad (13\ a)$$

$$\sigma''(\omega) = \frac{\sigma_b^2 C_{dl} \omega^\alpha \sin\left(\frac{\alpha\pi}{2}\right)}{\sigma_b^2 + 2\sigma_b C_{dl} \omega^\alpha \cos\left(\frac{\alpha\pi}{2}\right) + (C_{dl}\omega^\alpha)^2} + \omega C_b \quad (13\,b)$$

Now, to include the high-frequency JPL response in the real part of the conductivity and high-frequency response in the imaginary part of conductivity, these equations (14 a & b) are as given below

$$\begin{cases} \sigma'(\omega) = \sigma_b \left[1 + \left(\frac{\omega}{\omega_h}\right)^n\right] & (14\,a) \\ \quad\quad and \\ \sigma''(\omega) = A\omega^s & (14\,b) \end{cases}$$

Here, all parameters have their usual meaning as earlier and both 'n' & 's' have a value less than unity. For complete frequency window analysis, we replace the $\sigma_b$ in Eq. 13 a by Eq. 14 a and Eq. 13 b by Eq. 14 b and this final conclusive equation is used for stimulating the present system [72]. Where, $C_{dl}$ is frequency independent double layer capacitance, $\omega$ is the angular frequency, s & $\alpha$ are exponent terms with value <1 and $C_b$ is the bulk capacitance of solid polymer electrolyte.

Table 4. Comparison of fitted parameters for real part of conductivity for different SPEs at RT.

| Sample Code | $\sigma_b$ (×10⁻⁷ S cm⁻¹) | $\omega_h$ (× 10⁴) | $\alpha$ | n | $C_{dl}$ (μF) | $\tau_h$ (μs) |
|---|---|---|---|---|---|---|
| O/Li=0 | 0.17 | 71 | 0.68 | 0.67 | 1.34 | 1.39 |
| O/Li=14 | 3.96 | 295 | 0.39 | 0.69 | 9.43 | 0.33 |
| O/Li=16 | 5.53 | 2520 | 0.44 | 0.38 | 19.04 | 0.03 |
| O/Li=18 | 3.70 | 944 | 0.54 | 0.43 | 15.62 | 0.10 |
| O/Li=20 | 5.95 | 259 | 0.25 | 0.87 | 16.83 | 0.38 |

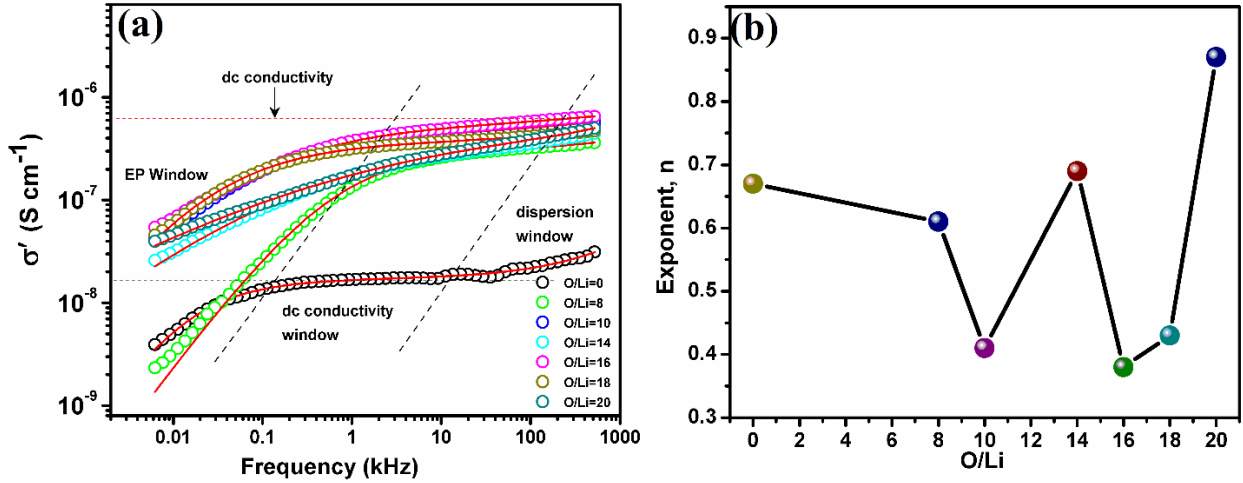

**Figure 10.** (a) Real part of complex conductivity (σ′) for polymer blend PEO/PAN+LiPF₆ based solid polymer electrolyte and (b) Variation of the power law exponent with salt concentration. Solid lines are the best fit to the experimental data.

Figure 10 shows the profile of the real part of conductivity and solid red lines are perfect corresponding fits. The simulated results are in absolute agreement with the experimental data points, and the fitting of the complete frequency window analysis will offer better Dielectric Parameters analysis than the JPL fit. Now, in the low-frequency region, the faster rate of ion jump from one coordinating site to another increases the relaxation time due to long-range ion transport. But, at the high-frequency inverse trend is observed and two competing processes occur one is unsuccessful hopping and another is successful hopping. The former one is said to happen when an ion jumps back to its initial position and later one is when the neighborhood ions become relaxed concerning to the ion's position (the ions stay in the new site) [73]. In the high-frequency region hopping frequency plays an active role in deciding the process of ion mechanism. Generally, for frequencies greater than the hopping frequencies, later one dominates and leads to the dispersion in the ac conductivity pattern [74]. The fitted parameters are listed in Table 4. The low value of both exponent parameter suggests that the investigated system is the purely ionic type. All parameters get changed with the addition of the salt in the polymer blend, and evidence the disorder produced in the polymer chain which reflects the enhanced polymer chain flexibility and ion dynamics. Also, it was revealed from the deep analysis of Table 4 that both hopping frequency and double layer capacitance increase with the addition of the salt and highest double layer capacitance value is obtained for the polymer electrolyte with highest ionic conductivity. Also, the decrease of hopping relaxation time indicates the rapid ion transport inside the polymer matrix which is in correlation with the impedance study. The lowest value of the relaxation time for the highest ionic conductive polymer electrolyte confirms the study in the previous sections and provide substantial evidence of the fast SSIC. Further, the comparative analysis will be done in the forthcoming section for an absolute understanding of the effect of salt in the polymer blend matrix.

*3.5.2. The imaginary part of complex conductivity*

As, complex conductivity comprises of two parts, real and imaginary parts of the conductivity. Figure 11 shows the frequency dependent imaginary part of conductivity and solid line in the plot are best-stimulated results. The fitted parameters are listed in Table 5. From the fitting results, it can be said that experimental data are in absolute agreement with the simulated data which provides consistency to the investigated system. For the better understating of the frequency dependent imaginary part of the conductivity, the results will be correlated with the real part of the conductivity where required. All plots depict the increase in the imaginary part of the conductivity. The careful investigation of the graph provides us two parameters one is maximum frequency ($\omega_{max}$) and another is onset frequency ($\omega_{on}$). The later one is the frequency on which the polarization starts and former one denotes the frequency at which maximum polarization is achieved [75-76]. For the onset frequency, all plots show a minimum in the $\sigma''$ which corresponds to the dispersive region in the plot of real part of conductivity with the frequency (right portion of the plot). So, with the decrease of frequency imaginary part of the conductivity decreases and after crossing the onset frequency $\sigma''$ again increases with frequency. A peak appears for the maximum $\sigma''$ on the frequency $\omega_{max}$ and at this frequency maximum polarization is achieved by the corresponding polymer electrolyte system.

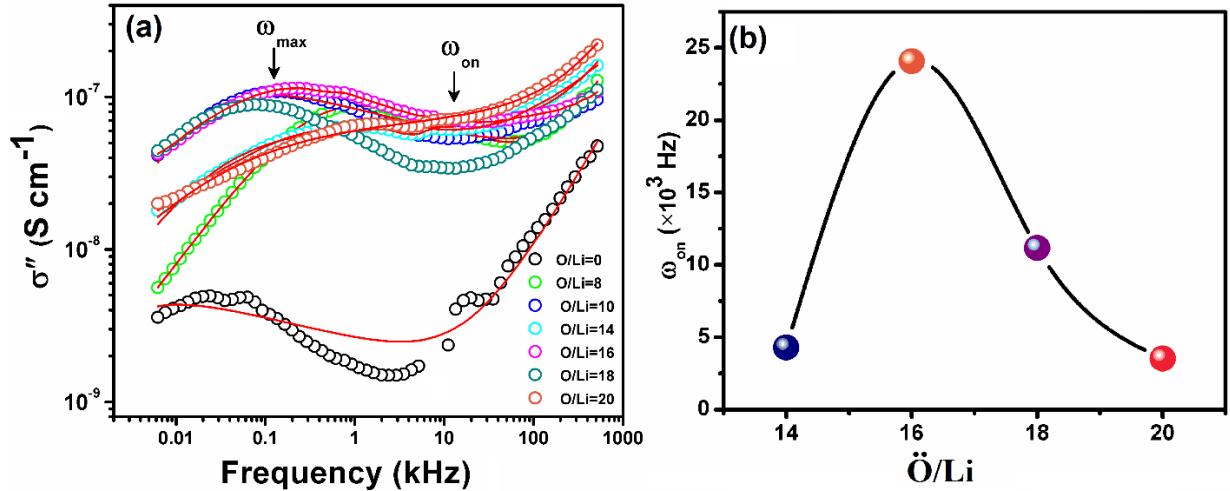

**Figure 11. (a)** Imaginary part of complex conductivity ($\sigma''$) for polymer blend PEO/PAN+LiPF$_6$ based solid polymer electrolyte and **(b)** Variation of onset frequency ($\omega_{on}$) against the salt concentration. Solid lines are the best fit to the experimental data.

This maximum frequency peak in the plot of the imaginary part of the conductivity with frequency corresponds to the decrease in the real part of ac conductivity. Also, the onset of EP or minimum in the plot of the complex conductivity ($\sigma''$) corresponds to the maxima in the tangent delta plot analysis as shown in Figure 12 (a & b) [77]. Then, after crossing the peak in the imaginary part of conductivity again the $\sigma''$ decreases. Figure 11 b shows the variation of onset polarization frequency with the salt content. Also, the onset frequency shifts toward the high frequency on the addition of salt in polymer blend and increased polarization region is obtained [78-79].

Table 5. Comparison of fitted parameters for the imaginary part of conductivity for different SPEs at RT.

| Sample Code | A (S cm$^{-1}$) | s | C$_{dl}$ (µF) | α | $C_h (\times 10^{-13}) F$ |
|---|---|---|---|---|---|
| O/Li=0 | 5.32×10$^{-10}$ | 0.14 | 1.34 | 0.97 | 0.96 |
| O/Li=14 | 3.07×10$^{-8}$ | 0.05 | 9.43 | 0.85 | 2.11 |
| O/Li=16 | 1.49×10$^{-7}$ | 0.13 | 19.04 | 0.82 | 1.39 |
| O/Li=18 | 8.07×10$^{-9}$ | 0.20 | 15.62 | 0.98 | 2.11 |
| O/Li=20 | 5.50×10$^{-8}$ | 0.04 | 16.83 | 0.67 | 2.72 |

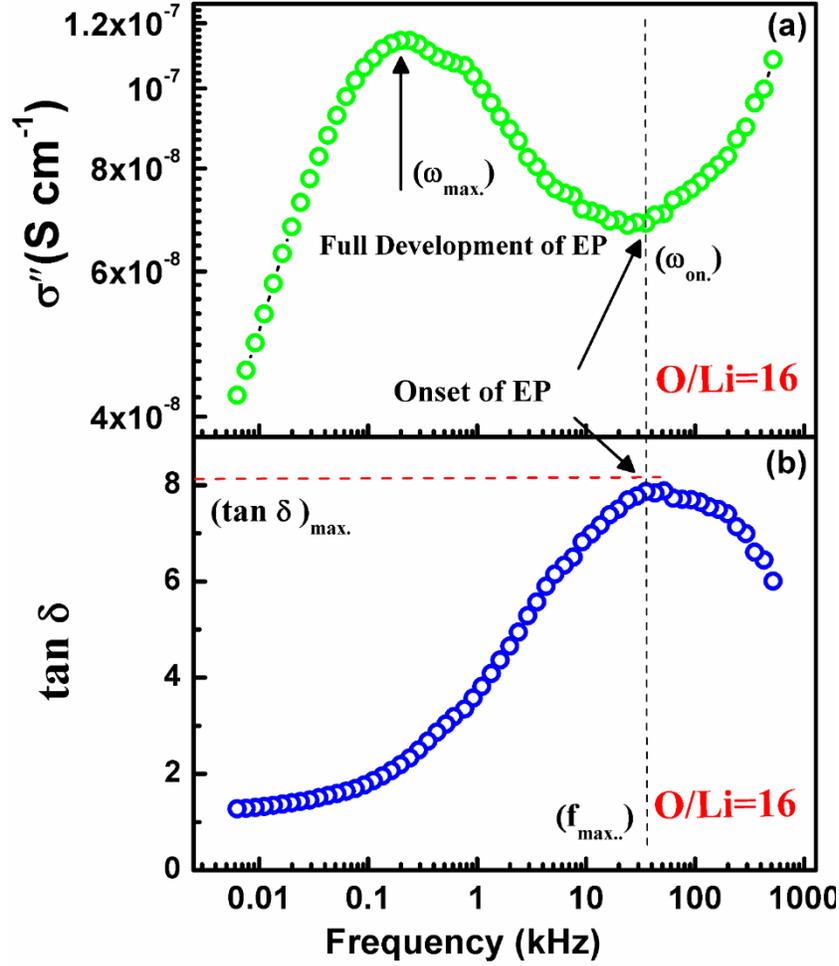

**Figure 12**. The imaginary part of complex conductivity and the tangent delta plot for PEO-PAN with O/Li=16.

### 3.6. Modulus Spectroscopy

The electric modulus study is extensively used to interpret the bulk dielectric behavior in case of the polymer electrolytes as a function of the frequency. In the Cole-Cole plot, the high value of both real and imaginary part of complex permittivity at the low-frequency window is due to electrode polarization (EP) event. The modulus formalism was used to separate the bulk relaxation phenomena from the ionic relaxation [58, 80]. Now, the dielectric data was transformed into the modulus data and complex modulus ($M^*$) can be expressed in relation with the complex dielectric permittivity ($\varepsilon^*$) by equation 15 a & b;

$$\begin{cases} M^* = M' + jM'' = \dfrac{1}{\varepsilon^*}\;;\; M' = \dfrac{\varepsilon'}{\varepsilon'^2 + \varepsilon''^2}\; and\; M'' = \dfrac{\varepsilon''}{\varepsilon'^2 + \varepsilon''^2} & (15\,a) \\ \quad\quad\quad\quad\quad\quad\quad\quad and \\ \quad M^* = j\omega C_o Z^* = \omega C_o Z'' + j\omega C_o Z' & (15\,b) \end{cases}$$

Here, $M'\;and\;M''$ are the real and imaginary part of the complex modulus ($M^*$), respectively. Figure 13 shows the plot of $M''$ vs. $M'$ for different salt concentration and all display the same nature of a deformed semicircle. This kind of deformed semi-circle evidences the broad relaxation processes and also verifies the broad relaxation peak observed in the tangent delta plot. The complete semi-circle is not obtained here because relaxation peak lies outside the measured frequency range.

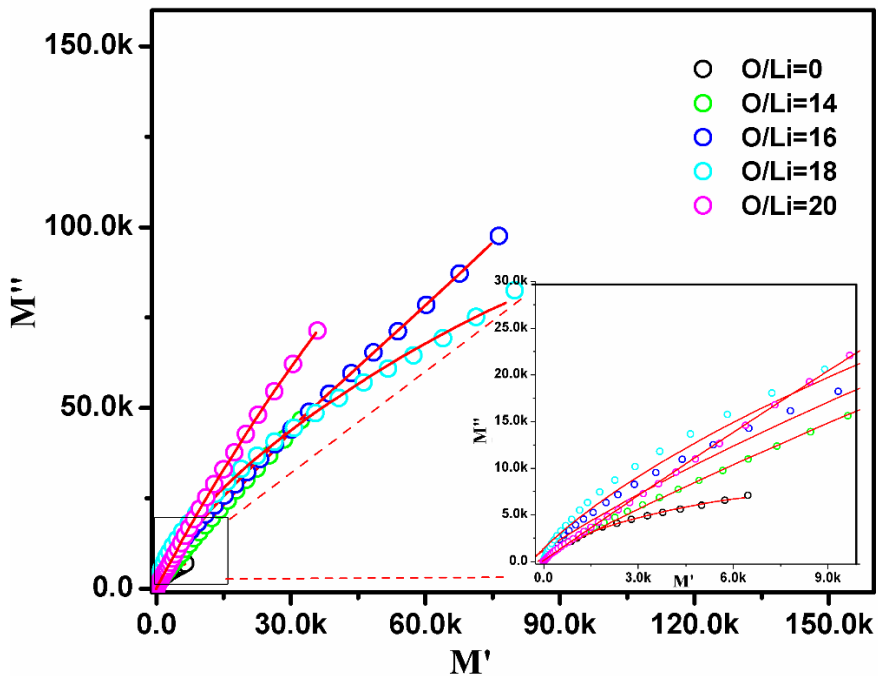

**Figure 13.** Argand plot of M" vs. M' for the polymer blend PEO/PAN+LiPF$_6$ based solid polymer electrolyte.

Figure 14 a & b depicts the frequency dependence of the real and imaginary parts of the electric modulus. In figure 14 a real modulus (M′) approaches to zero at low frequency and presence of long tail may be due to the electrode polarization effect which was associated with the large double layer capacitance. When the frequency increases then it is noticed that the dispersion comes into account and saturation in the real part of modulus is achieved at the very high frequency. This dispersion in the real part of the modulus corresponds to the peak in the imaginary part of the modulus spectra. This type of behavior was observed when there is less restoring force for mobile charges on the application of the field, and this evidences the long-range mobility of charge carriers along with the segmental relaxation. The region on the left of the peak indicates the long-range mobility of the ions while the right side of the peak is an indication of the bounded ions in the potential wells [81-86]. In the plot of the imaginary part of modulus against frequency (Figure 14 b) the relaxation peak is not observed for the polymer salt system and is not within the experimental frequency range. Same reports have been reported earlier also [87-88].

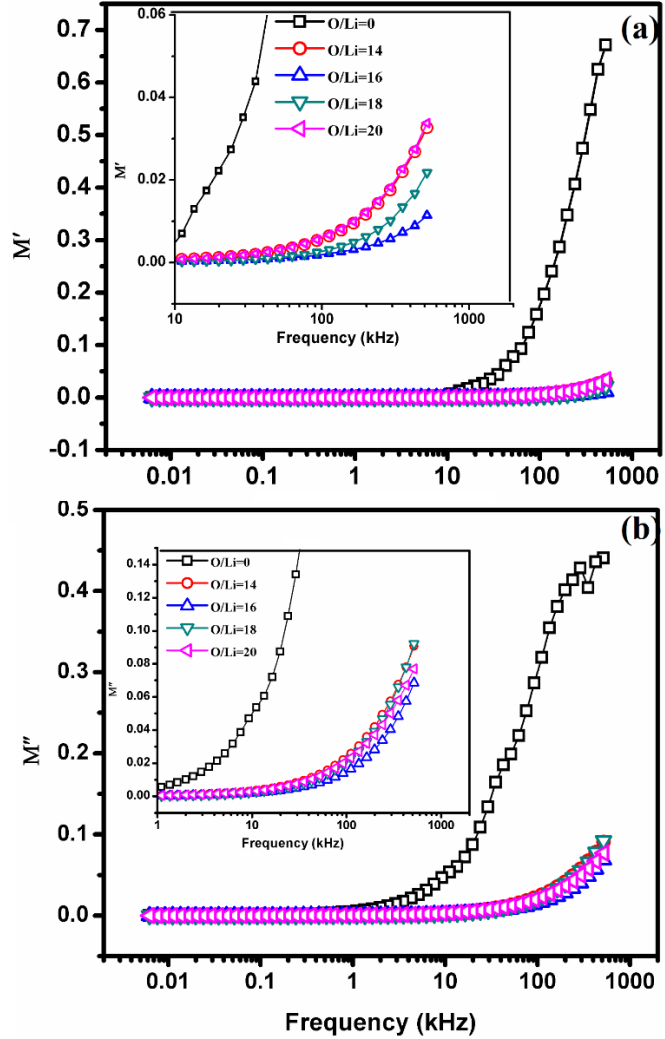

**Figure 14.** Frequency dependence of the (a) real part of modulus (M′) and (b) imaginary part of modulus (M″) for polymer blend PEO/PAN+LiPF$_6$ based solid polymer electrolyte.

This relaxation peak somewhere lies at a high frequency so, the relaxation time was obtained from the tangent delta analysis and a decrease in relaxation time ($\tau_M$) indicates that peak gets shifted toward higher frequency on the addition of salt in the polymer blend. The reduction in the value of the relaxation time indicates that the ion dynamics is improved on the addition of salt. After exploring the dielectric parameters now, the study was focused on built up of a correlation of the various dielectric parameters calculated recently.

### 3.7. Correlation between the ionic conductivity, double layer capacitance, dielectric strength and various relaxation times ($\tau_{\varepsilon'}, \tau_{\tan\delta}, \tau_M, \tau_m, \tau_h$)

Since, dielectric ion dynamics brief analysis gives in hand various parameters that are necessary for understanding the ion dynamics such as, dielectric strength, relaxation time. So, all concerned parameters are plotted together against the salt concentration for better visibility of the eye (Figure 15 a-h). The comparative plot together shows the one-to-one correspondence between ionic conductivity, double layer capacitance and the relaxation times. Note that the conductivity is directly dependent on the number of free charge carriers and dissociation of the salt within the polymer matrix. From FTIR deconvolution it was concluded that number of free charge carriers show maxima and ion pair minima for the O/Li=16. The highest ionic conductivity was achieved for the O/Li=16 and may be attributed to the complete dissociation of the salt as shown in Figure 15 a. Also, the double layer capacitance depends on the number of free charge carriers and that also showed the maxima for the same concentration (Figure 15 b). Dielectric strength (Δε) or the ionic polarization exhibits maxima for the same concentration and indicates better salt dissociation for this composition.

Now, as the cation migration in the investigated system occurs via the formation and breaking of coordinating sites in the polymer chain. This jump of the cation from one site to another takes some time and that depends on the segmental mobility of the polymer chain. In this regard, various relaxation times were calculated and it was clearly observed from the Figure 15 d-h that the all plots show minima for the O/Li=16 concentration. That strongly supports our approach of the ion transport via the segmental motion of polymer chain. In this article, we have also calculated the molecular relaxation time ($\tau_M$) that was associated with the intrinsic dipoles of the polymer chain. The molecular relaxation time also follows the same trend and is vital evidence of the enhanced transport properties. This correlation is also in absolute agreement with the Sigma representation analysis which shows the increase in the radius of the semi-circle, hence the decrease of the relaxation time. For the high concentration increase in the relaxation time is an indication of the delayed ion migration from one site to another which directly evidences the decrease of ion conductivity. So, from the whole broad discussion, it can be summarized that the one-to-one correspondence between all the parameters validates the genuineness of investigated polymer electrolyte system.

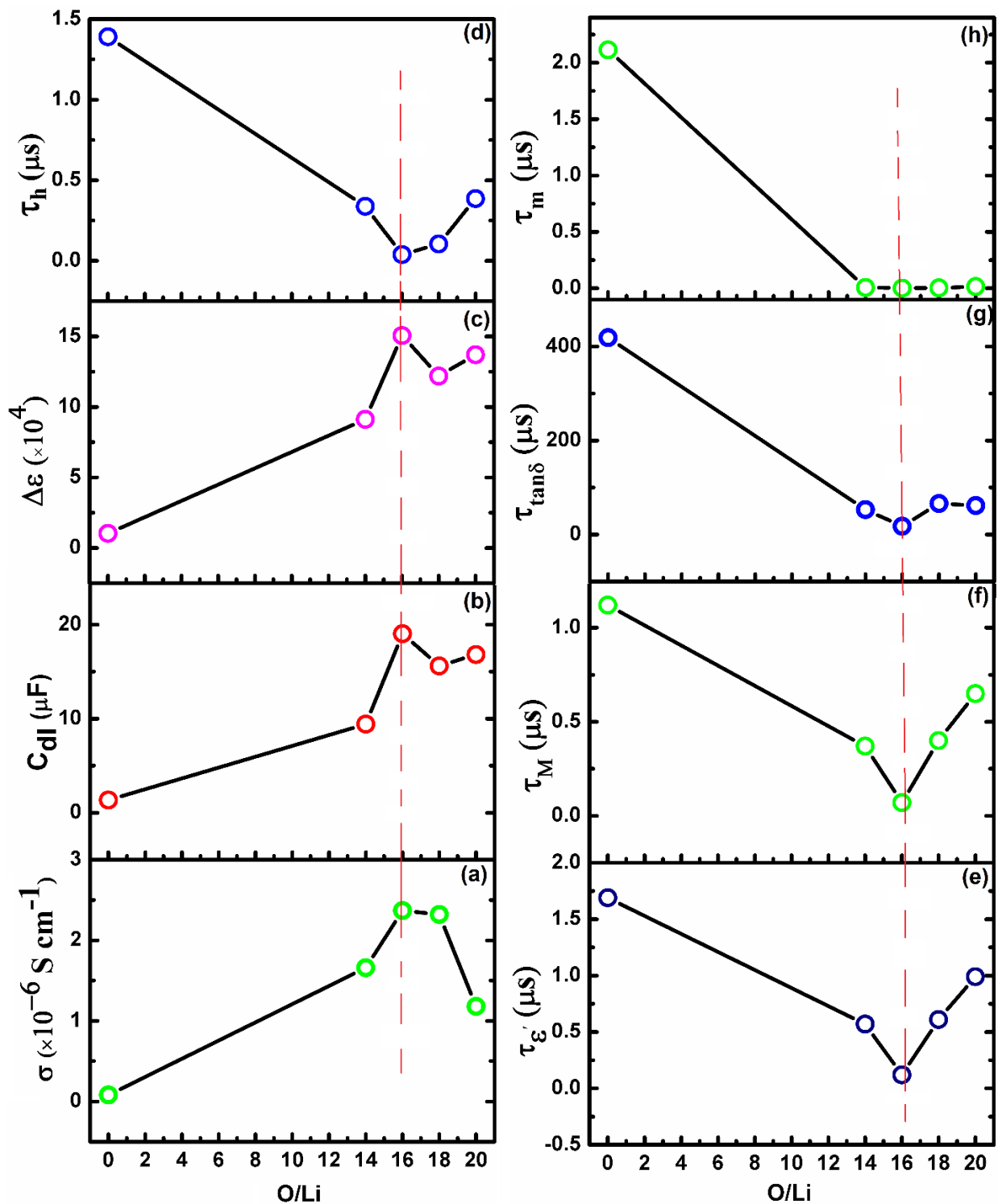

**Figure 15.** Plot of (a) ionic conductivity, (b) double layer capacitance, (c) dielectric strength and (d-h) various relaxation times ($\tau_{\varepsilon'}, \tau_{\tan\delta}, \tau_M, \tau_m, \tau_h$)) for the PEO/PAN blend and with O/Li=14, 16, 18, 20. Red line in the graph shown to the guide of the eye.

**3.8. Proposed Mechanism**

The above-mentioned experimental findings insight to propose a simplified model to explain the ion dynamics and transport properties of polymer electrolyte systems. Initially, both PEO and PAN interact via the hydrogen bonding as shown in Figure 16 a. When salt is added to the polymer matrix, then it gets dissociated into cations and anions. Two possibilities arise here for cation, (i) cation coordination with ether group of PEO, and (ii) cation coordination with nitrile group of PAN. As it was evidenced by the FTIR analysis that cation is going to coordinate with the ether group of PEO. Also, the electronegativity of ether group is more than nitrile group and evidences the feasible interaction of cation with the former one. So, the mechanism highlights the individual role of PEO polymer chain rather than both during explanation of interaction with salt. In figure 16 b, lithium get coordinated with the ether group of the polymer chain. Since lithium is Lewis acid and ether group is Lewis base, so the Lewis-acid-base interaction makes the interaction between them. Therefore, lithium gets dissociated, and the anion gets attached to the polymer chain. This dissociated cation ($Li^+$) will play a vital role in the transport. Now, in figure 16 c, it can be seen that at any instant lithium is coordinated with the four electron rich sites (ether group) of the polymer chain. Now, as the electric field is applied then lithium is triggered by the filed in its direction and it breaks the current coordination bond. Therefore, the new coordination bond formed with another site of the polymer chain. Thus, the triggering of the ion from one site to another is supported by the polymer chain, and its flexibility plays an active role in the ion transportation. But, it is well known that polymer chain flexibility is more when the amorphous content is enhanced. The improved amorphous content with the addition of the salt leads to the formation of favorable conduction paths supported by the backbone of the polymer chain and results in the enhancement of the ionic conductivity or the reduction of relaxation time. This quick jump from one coordinating site to another also depends upon the number of free charge carriers or the dielectric constant which decides the dissociation rate of the salt. In this whole process of transportation, since cation is the main dominating species, therefore, conduction contribution counted by only cation contribution. As, the anion is bulky in size, it is supposed to be attached to the backbone or the methyl group of the polymer chain. The proposed mechanism leads to better visibility in terms of enhancement in the mobility of the polymer chain, high ionic conductivity and reduced molecular relaxation time.

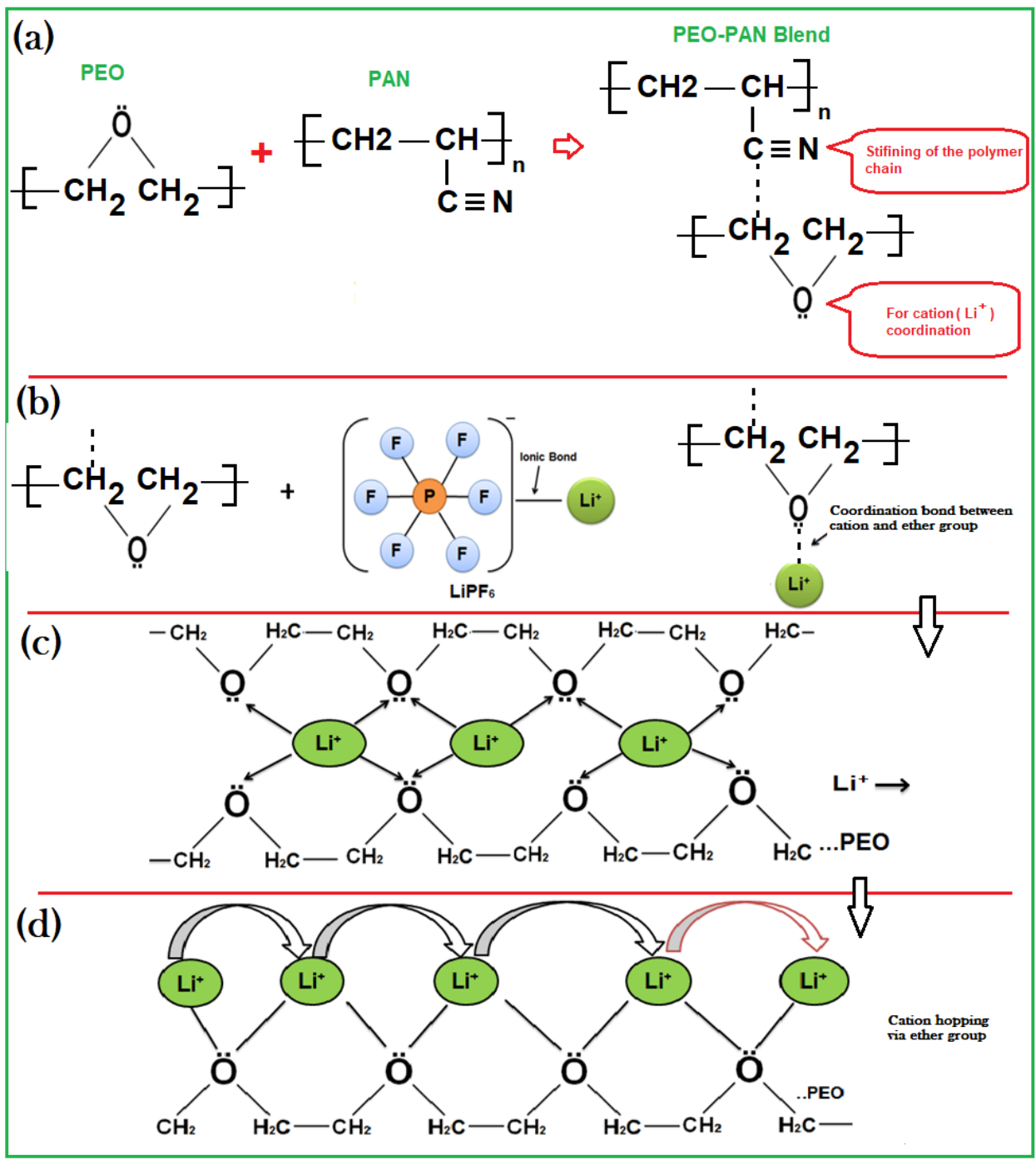

**Figure 16.** Proposed model for the lithium ion transport, (a) blend polymer electrolyte formation (b) Formation of coordination between the ether group of PEO and cation (Li$^+$), (c) Translational motion of cation via electron rich ether group of host polymer, (d) symbolic representation of the triggering of lithium ion from one coordinating site to another.

## 4. Conclusion

Blend Polymer (PEO/PAN) lithium ion conducting films have been prepared and microstructural, impedance and dielectric analysis have been studied in a systematic fashion. The complex permittivity, electrical conductivity, tangent-delta loss, and modulus formalism provide better insight of ion dynamics. From the FTIR analysis, the complex formation and presence of the ion-ion and polymer-ion interaction were confirmed. The deconvolution of FTIR provides us the free ion area and found maximum for the O/Li=16 concentration while for the higher salt concentration the ion pair area dominates leading to the decrease of the conductivity. Cole-Cole plots provided sufficient information such as the dielectric strength and dielectric permittivity at high frequency. The dielectric permittivity and dielectric loss show the decrease with the increase in the frequency. Also, the high dielectric constant was observed for the O/Li=16 concentration which means the highest ionic conductivity. Tangent delta loss relaxation peak shows a shift toward the high-frequency window which indicates the faster ion migration or enhanced conductivity. The sigma representation ($\sigma''$ $vs.$ $\sigma'$) shows the increase in the radius of the semi-circle which indicates the faster ion dynamic and is beneficial for the fast solid state ionic conductor. The real and imaginary part of the complex conductivity were fitted in the complete frequency range and shows absolute agreement with the experimental data. Finally, the correlation between the various dielectric parameters is displayed and show one-to-one correspondence with all other results. We may thus anticipate that the observed remarkable correlation between the various fitting parameters allows us to further go in detail to understand the ion dynamics in case of polymer electrolytes and Sigma representation may provide more crucial information about the relaxation dynamics.

**Acknowledgment**

One of the authors acknowledges with thanks for financial support from CUPB and partial funding from UGC Startup Grant (GP-41).

**References**

1. H. Zhang, C. Li, M. Piszcz, E. Coya, T. Rojo, L. M. Rodriguez-Martinez, M. Armand, and Z. Zhou, Chemical Society Reviews **46**, 797 (2017).
2. M. Armand and J. M. Tarascon, Nature **451**, 652 (2008).
3. A. Arya and A. L. Sharma, Ionics **23**, 497 (2017).
4. G. Zhou, F. Li, and H. M. Cheng, Energy and Environmental Science **7**, 1307 (2014).
5. D. Zhao, S. Ge, E. Senses, P. Akcora, J. Jestin, and S. K. Kumar, Macromolecules **48**, 5433 (2015).
6. J. Xi, X. Qiu, J. Li, X. Tang, W. Zhu, and L. Chen, Journal of Power Sources **157**, 501 (2006).
7. K. Deng, S. Wang, S. Ren, D. Han, M. Xiao, and Y. Meng, Journal of Power Sources **360**, 98 (2017).
8. J. Zhang, N. Zhao, M. Zhang, Y. Li, P. K. Chu, X. Guo, Z. Di, X. Wang, and H. Li, Nano Energy **28**, 447 (2016).
9. A. L. Sharma, N. Shukla, and A. K. Thakur, Journal of Polymer Science Part B: Polymer Physics **46**, 2577 (2008).
10. M. Y. Yoon, S. K. Hong, H. J. Hwang, Ceram. Int. **39**, 9659 (2013).
11. G. S. MacGlashan and Y. G. Andreev, Nature **398**, 792 (1999).
12. V. Di Noto, S. Lavina, G. A. Giffin, E. Negro, and B. Scrosati, Electrochimica Acta **57**, 4 (2011).
13. P. A. R. D. Jayathilaka, M. A. K. L. Dissanayake, I. Albinsson, and B. E. Mellander, Electrochimica Acta **47**, 3257 (2002).
14. C. Bhatt, R. Swaroop, A. Arya, A. L. Sharma, Journal of Materials Science and Engineering B **5**, 418 (2015).
15. A. Arya and A. L. Sharma, Journal of Physics D: Applied Physics **51**, 045504 (2018).

16. A. Arya, M. Sadiq, and A. L. Sharma, Ionics **24**, 2295 (2018).
17. Z. Zainuddin, D. Hambali, I. Supa'at, and Z. Osman, Ionics **23**, 265 (2017).
18. A. L. Sharma and A. K. Thakur, Ionics **19**, 795 (2013).
19. J. Xi, X. Qiu, J. Li, X. Tang, W. Zhu, and L. Chen, Journal of Power Sources **157**, 501 (2006).
20. V. B. Achari, T. J. R. Reddy, A. K. Sharma, V. N. Rao Ionics **13**, 349 (2017).
21. A. Arya, A. L. Sharma, Journal of Solid State Electrochemistry 1-21 (2018). DOI: 10.1007/s10008-018-3965-4.
22. J. Cao, L. Wang, X. He, M. Fang, J. Gao, J. Li, L. Deng, H. Chen, G. Tian, J. Wang, and S. Fan, Journal of Materials Chemistry A **1**, 5955 (2013).
23. Y. Zhao, C. Wu, G. Peng, X. Chen, X. Yao, Y. Bai, F. Wu, S. Chen, and X. Xu, Journal of Power Sources **301**, 47 (2016).
24. M. B. Armand, Annual Review of Materials Science **16**, 245 (1986).
25. B. Scrosati, F. Croce, and S. Panero, Journal of Power Sources **100**, 93 (2001).
26. A. Arya, A. L. Sharma, Journal of Physics D: Applied Physics **50**, 443002 (2017).
27. K. Kesavan, C. M. Mathew, and S. Rajendran, Chinese Chemical Letters **25**, 1428 (2014).
28. A. Arya, A. L. Sharma, Applied Science Letters **2**, 72 (2016).
29. A. L. Sharma, A. K. Thakur, J Mater Sci. **46**, 1916 (2011).
30. L. N. Sim, F. C. Sentanin, A. Pawlicka, R. Yahya, and A. K. Arof, Electrochimica Acta **229**, 22 (2017).
31. R. Younesi, G. M. Veith, P. Johansson, K. Edstrom, and T. Vegge, Energy & Environmental Science **8**, 1905 (2015).
32. A. Arya, A. L Sharma, S. Sharma, M. Sadiq, Journal of Integrated Science and Technology **4**, 17 (2016).
33. H. Ohno, M. Yoshizawa, W. Ogihara, Electrochimica Acta **48**, 2079 (2003).
34. L. Fan, Z. Dang, C. W. Nan, M. Li, Electrochimica Acta **48**, 205 (2002).
35. P. S. Anantha and K. Hariharan, Materials Science and Engineering B: Solid-State Materials for Advanced Technology **121**, 12 (2005).
36. P. S. Anantha and K. Hariharan, Journal of Physics and Chemistry of Solids **64**, 1131 (2003).
37. S. Das and A. Ghosh, Electrochimica Acta **171**, 59 (2015).
38. A. Karmakar, A. Ghosh, Physical Review E **84**, 051802 (2011).
39. K. Kesavan, C. M. Mathew, S. Rajendran, and M. Ulaganathan, Materials Science and Engineering: B **184**, 26 (2014).
40. B. Jinisha, K. M. Anilkumarm M. Manoj, V. S. Pradeep, S. Jayalekshmi, Electrochimica Acta **235**, 210 (2017).
41. N. Bar, P. Basak, and Y. Tsur, Physical Chemistry Chemical Physics **19**, 14615 (2017).
42. S. Selvasekarapandian, R. Baskaran, O. Kamishima, J. Kawamura, T. Hattori, Spectrochimica Acta Part A: Molecular and Biomolecular Spectroscopy **65**, 1234 (2006).
43. S. K. Chaurasia, R. K. Singh, S. Chandra, Journal of Raman Spectroscopy **42**, 2168 (2011).
44. R. Tang, C. Jiang, W. Qian, J. Jian, X. Zhang, H. Wang, and H. Yang, Scientific Reports **5**, 13645 (2015).
45. A. K. Jonscher, Dielectric Relaxation in Solids, Chelsea Dielectric, London, 1983.
46. A. Sharma and A. Thakur, Applied Polymer Science **118**, 2743 (2010).
47. H. E. Atyia, N. A. Hegab, Optik-International Journal for Light and Electron Optics **127**, 6232 (2016).
48. Ahmad, Z. (2012). Polymer dielectric materials. In Dielectric material. InTech.

49. J. G. Powles, Proceedings of the Physical Society Section B **64**, 81 (1951).
50. F. Salman, R. Khalil, H. Hazaa, Adv. J. Phys. Sc. 3, 1 (2014).
51. K. S. Cole, J. Chem. Phys. **9**, 341 (1940).
52. A. Arya, S. Sharma, A. L. Sharma, D. Kumar, M. Sadiq, Asian J Eng Appl Technol **5**, 4 (2016).
53. W. Cao, R. Gerhardt, Solid State Ionics **42**, 213 (1990).
54. A. L. Sharma, A. K. Thakur, Ionics **21**, 1561 (2015).
55. G. Govindaraj, N. Baskaran, K. Shahi, and P. Monoravi, Solid State Ionics **76**, 47 (1995).
56. E. M. Masoud, A.-A. El-Bellihi, W. A. Bayoumy, and M. A. Mousa, Journal of Alloys and Compounds **575**, 223 (2013).
57. N. Tripathi, A. K. Thakur, A. Shukla, and D. T. Marx, Polymer Engineering and Science **58**, 220 (2018).
58. N. Chilaka, S. Ghosh, Electrochimica Acta **134**, 232 (2014).
59. M. Ravi, Y. Pavani, K. Kiran Kumar, S. Bhavani, A. K. Sharma, and V. V. R. Narasimha Rao, Materials Chemistry and Physics **130**, 442 (2011).
60. R. J. Singh, Solid State Physics, 2012, Dorling Kindersley (India).
61. S. Chopra, S. Sharma, T. C. Goel, and R. G. Mendiratta, Solid State Communications **127**, 299 (2003).
62. R. J. Sengwa, S. Choudhary, Journal of Alloys and Compounds **701**, 652 (2017).
63. R. M. Hill, A. K. Jonscher, Contemp Phys. **24**, 75 (1983).
64. K. Nakamura, T. Saiwaki, K. Fukao, Macromolecules **43**, 6092 (2010).
65. Y. Wang, C. N. Sun, F. Fan, J. R. Sangoro, M. B. Berman, S. G. Greenbaum, T. A. Zawodzinski, and A. P. Sokolov, Physical Review E - Statistical, Nonlinear, and Soft Matter Physics **87**, 042308 (2013).
66. Y. Z. Wei and S. Sridhar, The Journal of Chemical Physics **99**, 3119 (1993).
67. A. L. Sharma and A. K. Thakur, Ionics **17**, 135 (2010).
68. M. Sadiq, A. Arya, A. L. Sharma, Recent Trends in Materials and Devices. Recent Trends in Materials and Devices, Springer International Publishing **178**, 389 (2017).
69. S. Nasri, A. L. Ben Hafsia, M. Tabellout, and M. Megdiche, Rsc Advances **6**, 76659 (2016).
70. A. K. Jonscher, Journal of Materials Science **13**, 553 (1978).
71. A. L. Sharma, A. K. Thakur, Ionics **16**, 339 (2010).
72. A. Roy, B. Dutta, S. Bhattacharya, RSC Advances **6**, 65434 (2016).
73. N. Shukla, A. K. Thakur, A. Shukla, and D. T. Marx, International Journal of Electrochemical Science **9**, 7644 (2014).
74. S. Choudhary, R. J. Sengwa, J. Appl. Polym. Sci. **132**, 41311 (2015).
75. Y. Wang, C. N. Sun, F. Fan, J. R. Sangoro, M. B. Berman, S. G. Greenbaum, T. A. Zawodzinski, and A. P. Sokolov, Physical Review E - Statistical, Nonlinear, and Soft Matter Physics **87**, 042308 (2013).
76. A. García-Bernabé, A. Rivera, A. Granados, S. V. Luis, and V. Compañ, Electrochimica Acta **213**, 887 (2016).
77. R. J. Klein, S. Zhang, S. Dou, B. H. Jones, R. H. Colby, and J. Runt, Journal of Chemical Physics **124**, (2006).
78. I. Fuentes, A. Andrio, F. Teixidor, C. Viñas, and V. Compañ, Physical Chemistry Chemical Physics **19**, 15177 (2017).
79. A. A. Khamzin, I. I. Popov, and R. R. Nigmatullin, Physical Review E - Statistical, Nonlinear, and Soft Matter Physics **89**, 032303 (2014).

80. M. Kumar, T. Tiwari, and N. Srivastava, Carbohydrate Polymers **88**, 54 (2012).
81. T. Mohamed Ali, N. Padmanathan, and S. Selladurai, Ionics **19**, 1115 (2013).
82. F. S. Howell, R. A. Bose, P. B. Macedo, and C. T. Moynihan, Journal of Physical Chemistry **78**, 639 (1974).
83. A. Arya and A. L. Sharma, Journal of Physics Condensed Matter **30**, 165402 (2018).
84. S. Das, A. Ghosh, The Journal of Physical Chemistry B **121**, 5422 (2017).
85. R. Arunkumar, R. S. Babu, and M. Usha Rani, Journal of Materials Science: Materials in Electronics **28**, 3309 (2017).
86. S. Choudhary, Journal of Materials Science: Materials in Electronics 1-18 (2018). DOI:10.1007/s10854-018-9116-y.
87. S. Choudhary and R. J. Sengwa, Materials Chemistry and Physics **142**, 172 (2013).
88. S. Choudhary, R. J. Sengwa, Polymer Bulletin, **72**, 2591 (2015).